\documentclass[%        Class options:
aps,%                   American Physical Society
prd,%                   Physical Review D
showpacs,%              Displays PACS after abstract
preprint,%              Preprint layout
tightenlines,%          Single spaced lines
superscriptaddress,%   Authors' addresses linked with superscripts
nofootinbib,%           Does not treat footnotes as references
floatfix]%              Fixes float errors
{revtex4}%              REVTEX 4 Package used
\usepackage{graphicx,%  Default Latex 2eps package for embedding figures
longtable%             Useful for long table
}
\begin{document}
%
%%%%%%%%%%%%%%% NEW COMMANDS %%%%%%%%%%%%%%%%%%%%%%%%%%%%%%%%%%%%%%%%%%%%%
\newcommand{\Lesssim}{\stackrel{\textstyle <}{\raisebox{-.6ex}{$\sim$}}}
\newcommand{\Gtrsim}{\stackrel{\textstyle >}{\raisebox{-.6ex}{$\sim$}}}
%%%%%%%%%%%%%%%%%%%%%%%%%%%%%%%%%%%%%%%%%%%%%%%%%%%%%%%%%%%%%%%%%%%%%%%%%%

%===========================================================================
\title{Addendum to:\\
Solar neutrino oscillation parameters\\ after first KamLAND
results}
%===========================================================================
%
\author{        G.L.~Fogli}
\affiliation{   Dipartimento di Fisica
                and Sezione INFN di Bari\\
                Via Amendola 173, 70126 Bari, Italy\\}
\author{        E.~Lisi}
\affiliation{   Dipartimento di Fisica
                and Sezione INFN di Bari\\
                Via Amendola 173, 70126 Bari, Italy\\}
\author{        A.~Marrone}
\affiliation{   Dipartimento di Fisica
                and Sezione INFN di Bari\\
                Via Amendola 173, 70126 Bari, Italy\\}
\author{        D.~Montanino}
\affiliation{   Dipartimento di Scienza dei Materiali
                and Sezione INFN di Lecce\\
                Via Arnesano, 73100 Lecce, Italy\\}
\author{        A.~Palazzo}
\affiliation{   Dipartimento di Fisica
                and Sezione INFN di Bari\\
                Via Amendola 173, 70126 Bari, Italy\\}
\author{        A.M.~Rotunno}
\affiliation{   Dipartimento di Fisica
                and Sezione INFN di Bari\\
                Via Amendola 173, 70126 Bari, Italy\\}
\begin{abstract}%...........................................................
In a previous paper \protect\cite{Ours}, we presented a
three-flavour oscillation analysis of the solar neutrino
measurements and of the first data from the KamLAND experiment, in
terms of the relevant mass-mixing parameters $(\delta
m^2,\theta_{12},\theta_{13})$. The analysis, performed by
including the terrestrial neutrino constraints coming from the
CHOOZ (reactor), KEK-to-Kamioka (K2K, accelerator) and
Super-Kamiokande (SK, atmospheric) experiments, provided a
stringent upper limit on $\theta_{13}$, namely,
$\sin^2\theta_{13}<0.05$ at $3\sigma$. We reexamine such upper
bound in the light of a recent (although preliminary) reanalysis
of atmospheric neutrino data performed by the SK collaboration,
which seems to shift the preferred value of the largest neutrino
square mass difference $\Delta m^2$ downwards. By taking the
results of the  SK official reanalysis at face value, and by
repeating the analysis in \protect\cite{Ours} with such a new
input, we find that the upper bound on $\theta_{13}$ is somewhat
relaxed: $\sin^2\theta_{13}<0.067$ at $3\sigma$. Related
phenomenological issues are briefly discussed.
\end{abstract}%.............................................................
\medskip
 \maketitle

%%%%%%%%%%%%%%%%%%%%%%%%%%%%%%%%%%%%%%%%%%%%%%%%%%%%%%%%%%%%%%%%%%%%%%%
In a previous paper \cite{Ours},  we presented a three-flavor
oscillation analysis of data from KamLAND \cite{KamL} and solar
\cite{Sola} neutrino experiments, in terms of the three
mass-mixing parameters relevant for this data set, namely, the
smallest square mass difference ($\delta m^2=m^2_2-m^2_1$), and
the two mixing angles between the first and the other two neutrino
generations ($\theta_{12}$ and $\theta_{13}$ in standard notation
\cite{PDGr}). The analysis included also the constraints coming
from the following terrestrial neutrino experiments: CHOOZ
(reactor) \cite{CHOO}, KEK-to-Kamioka (K2K, accelerator)
\cite{K2Ks}, and Super-Kamiokande (SK, atmospheric) \cite{SKat}.
In particular, an approximate SK+K2K data combination \cite{Lett}
was used to constrain the largest neutrino square mass difference:
$\Delta m^2=(2.7\pm0.4)\times10^{-3}$ eV$^2$ ($1\sigma$)
\cite{Lett}. The parameter $\Delta m^2$ was then marginalized away
in the $3\nu$ analysis of the CHOOZ spectral data
\cite{Ours,Lett}, which depends on the four parameters ($\Delta
m^2,\delta m^2,\sin^2\theta_{12},\sin^2\theta_{13}$) \cite{Pala}.

Our summary of the constraints on the $3\nu$ parameters ($\delta
m^2,\sin^2\theta_{12},\sin^2\theta_{23}$) was given in Fig.~9 of
\cite{Ours}, in terms of the projected $\Delta\chi^2$ functions of
the global (solar+terrestrial) data fit. The same functions are
reported, for the sake of completeness, in Fig.~1 (solid curves).
In particular, the following upper bound was obtained on the
mixing angle $\theta_{13}$ \cite{Ours}):
%..................................................................
\begin{equation}
\label{sin2old} \sin^2\theta_{13} < 0.05\ (3\sigma, \mathrm{\ all\
data})\ .
\end{equation}
%...................................................................

After the work \cite{Ours}, a thorough analysis \cite{Marr} of the
first K2K spectral data \cite{K2Ks} has provided, in combination
with SK atmospheric neutrino data \cite{SKat}, a more reliable and
accurate estimate of the $\Delta m^2$ range \cite{Marr}:
%..................................................................
\begin{equation}
\label{Deltam2old} \Delta m^2 = (2.6 \pm 0.4) \times
10^{-3}\mathrm{\ eV}^2\ (1\sigma, \mathrm{\ SK+K2K})\ .
\end{equation}
%...................................................................
A very similar range has been found in a recent, independent
analysis of SK+K2K data \cite{Conc}. However, the mere decrease of
the $\Delta m^2$ best-fit value from $2.7$ \cite{Lett} to 2.6
\cite{Marr,Conc} (in units of $10^{-3}$ eV$^2$) does not induce
any perceptible change in the results summarized in Fig.~9 of
\cite{Ours}. A nonnegligible change can be instead induced by a
more substantial decrease of $\Delta m^2$, as possibly indicated
by a new SK data reanalysis \cite{EPSC}.

The SK Collaboration has recently presented the preliminary
results of a global reanalysis of the previous atmospheric
neutrino data (no new data included), which incorporates
improvements or changes of various basic ingredients, such as the
neutrino interaction simulator, the inner and outer detector
simulators, the data reduction process, the event reconstruction
algorithm, and the input atmospheric neutrino fluxes \cite{EPSC}.
It is claimed that each change slightly shifts the $\Delta m^2$
allowed region to lower values, the final best-fit value being
$\Delta m^2=2\times 10^{-3}$ eV$^2$ \cite{EPSC}, i.e., $1.5\sigma$
below the central value in Eq.~(\ref{Deltam2old}).

At present, we cannot recover from \cite{EPSC} enough information
to implement the above variations in a thorough, {\em ab initio\/}
analysis of the SK atmospheric data, and thus we cannot
independently check the above claim. However, it is tempting to
study, at least in a first approximation, the implications of
relatively low $\Delta m^2$ values \cite{EPSC} on our previous
$3\nu$ analysis \cite{Ours}. In particular, it is immediate to
recognize \cite{LBLN} that a significant downward shift of $\Delta
m^2$ weakens the upper bound on $\sin^2\theta_{13}$ coming from
the CHOOZ data \cite{CHOO}. Since the parameter
$\sin^2\theta_{13}$ has an enormous impact on basically all
aspects of current and future neutrino phenomenology, we think it
useful  to reexamine its previous probability distribution
\cite{Ours} in the light of the preliminary SK revised results
\cite{EPSC}.

To this purpose, we derive an approximate
$\chi^2_{\mathrm{SK}'}(\Delta m^2)$ function through graphical
reduction of the SK mass-mixing parameter fit in \cite{EPSC}, and
combine it with our (unaltered) $\chi^2_\mathrm{K2K}(\Delta m^2)$
function from the K2K spectral analysis in \cite{Marr}. We obtain
then the following ``revised'' estimate for $\Delta m^2$,
%..................................................................
\begin{equation}
\label{Deltam2new}   \Delta m^2 = (2.0^{+0.4}_{-0.3}) \times
10^{-3}\mathrm{\ eV}^2\ (1\sigma, \mathrm{\ SK}'+\mathrm{K2K})\ ,
\end{equation}
%...................................................................
where the errors, although asymmetric, turn out to scale linearly
up to $\sim\!3\sigma$.%
%-------------------------------------------
\footnote{In other words, the function $\chi^2_\mathrm{SK'+K2K}$
is well approximated by two half-parabolae. The SK$'$ label
indicates that we are using here the ``revised'' Super-Kamiokande
results from \protect\cite{EPSC}.}
%---------------------------------------------------------
Notice that the above estimate is compatible  with the $\Delta
m^2$ ranges independently preferred by the final analyses of the
MACRO \cite{MACR} and Soudan~2 \cite{SOUD} atmospheric neutrino
experiments.

Assuming the $\Delta m^2$ input from Eq.~(\ref{Deltam2new}), we
proceed to perform the global $3\nu$ fit as in \cite{Ours}, all
other phenomenological inputs being unchanged. The results are
shown in Fig.~1 (dotted curves).  As compared with the previous
results \cite{Ours} (solid curves), no noticeable change is seen
in the fit to the leading solar $\nu$ parameters $\delta m^2$ and
$\sin^2\theta_{12}$ (left and middle panel in Fig.~1). The
constraints on $\sin^2\theta_{13}$ are instead somewhat relaxed,
as expected. In particular, the intercept of the line at $\Delta
\chi^2=9$ with the dotted curve in the right panel of Fig.~1
provides the ``revised'' $3\sigma$ upper bound on
$\sin^2\theta_{13}$,
%..................................................................
\begin{equation}
\label{sin2new} \sin^2\theta_{13} < 0.067\ (3\sigma, \mathrm{\
all\ data})\ ,
\end{equation}
%...................................................................
to be compared with the previous one in Eq.~(\ref{sin2old}).
Essentially, the main effect of the ``revised'' SK atmospheric
neutrino results \cite{EPSC} consists in enlarging the upper
bounds on $\sin^2\theta_{13}$ by a factor $\sim\! 1.3$ at any C.L.
This is the main result of our Addendum.

We conclude with a few qualitative comments on possible
phenomenological implications of the relatively ``low'' $\Delta
m^2$ in Eq.~(\ref{Deltam2new}) and of the relatively ``weak''
upper bound on $\sin^2\theta_{13}$ in Eq.~(\ref{sin2new}).
Concerning long-baseline accelerator experiment, an increase in
the upper limit on $\sin^2\theta_{13}$ may, in general, enlarge
the discovery potential in the $\nu_\mu\to\nu_e$ channel. On the
other hand, low values of $\Delta m^2$ imply small
$\nu_\mu\to\nu_\tau$ appearance rates [$\propto(\Delta m^2)^2$] in
the same class of experiments. Weakening the bounds on
$\sin^2\theta_{13}$ might also weaken future KamLAND limits on
$\theta_{12}$ (when they will become competitive with solar
neutrino limits), since variations of $\sin^2\theta_{12}$ can be
partly traded for variations of $\sin^2\theta_{13}$ (both
affecting the KamLAND event rate in a similar way). Uncertainties
on the mixing angles may then affect other observables, e.g., the
effective Majorana mass in neutrinoless double beta decay.
Concerning atmospheric neutrinos, low values of $\Delta m^2$ and
relatively high values of $\sin^2\theta_{13}$ can make subleading
$3\nu$ effects somewhat more important. In particular, if the
best-fit value of $\delta m^2$ in Fig.~1 would increase with
future KamLAND data, then atmospheric neutrino analyses at zeroth
order in $\delta m^2/\Delta m^2$ might need an upgrade to include
higher-order effects. Of course, none of the above  effects can be
large enough to change significantly the overall picture of the
$3\nu$ oscillation phenomenology. However, current or predicted
ranges for several parameters and observables might need small
readjustments, should the revised SK atmospheric neutrino analysis
in \cite{EPSC} and the $\Delta m^2$ estimate in
Eq.~(\ref{Deltam2new}) be basically confirmed by more detailed
studies: The ``revised'' upper limit on $\sin^2\theta_{13}$ in
Eq.~(\ref{sin2new}) is just one relevant example.

\acknowledgments

This work was in part supported by the Italian {\em Ministero
dell'Istruzione, Universit\`a e Ricerca\/} (MIUR) and {\em
Istituto Nazionale di Fisica Nucleare\/} (INFN) within the
``Astroparticle Physics'' research project.

%%%%%%%%%%%%%%%%%%%%%%%%%%%%%%%%%%%%%%%%%%%%%%%%%%%%%%%%%%%%%%%%%%

%%%%%%%%%%%%%%%%%%%%%%%%%%%%%%%%%%%%%%%%%%%%%%%%%%%%%%%%%%%%%%%%%%%%%%%%%%%%%%%%%%%
%---------------------------------------------------------------------------
\begin{figure}
\vspace*{0cm}\hspace*{-8mm}
\includegraphics[scale=0.86, bb= 100 100 500 720]{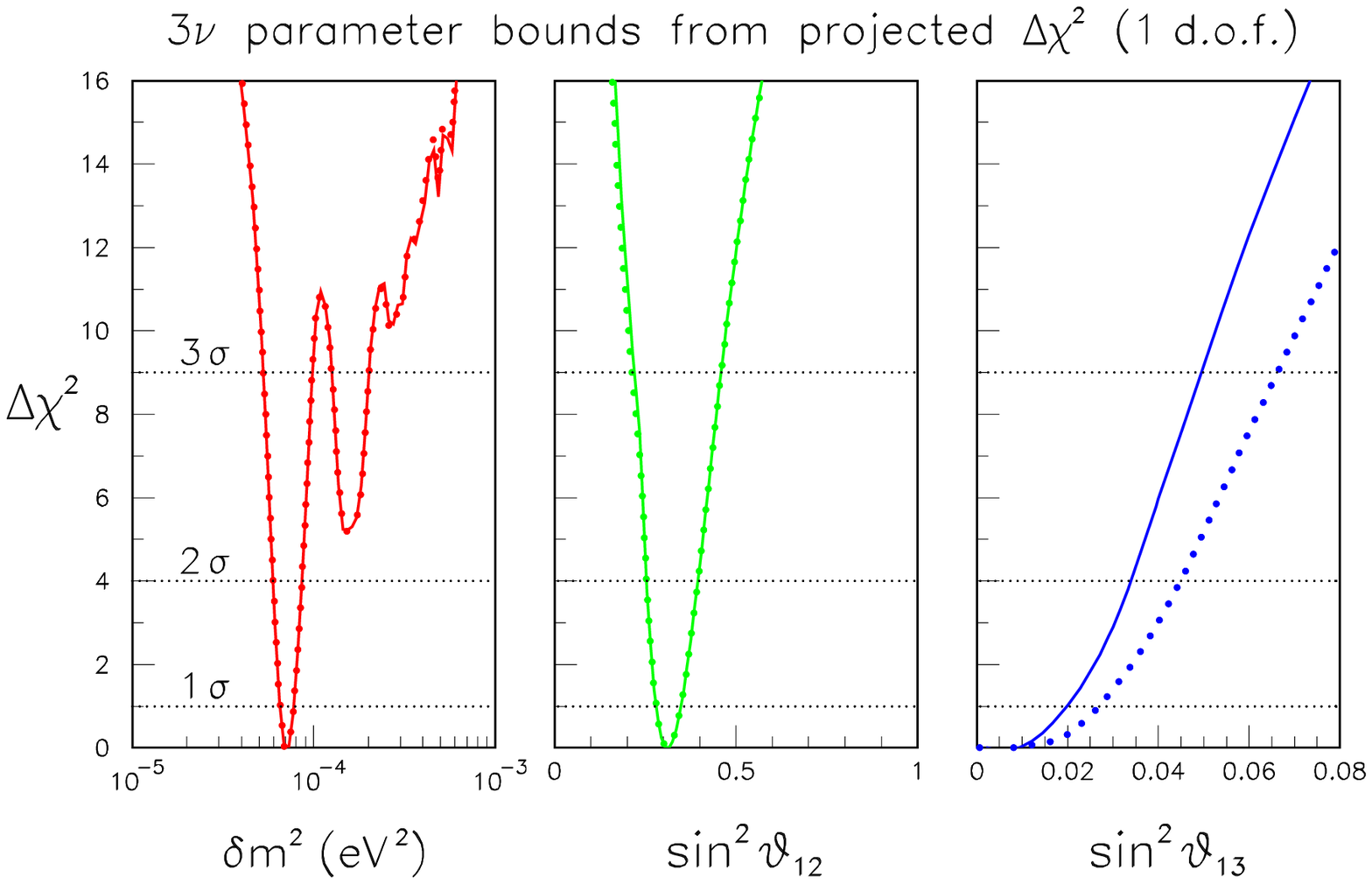}
\vspace*{-1.5cm}  \caption{\label{fig01} Three-flavor neutrino
oscillations: Projections of the global $\Delta \chi^2$ functions
onto each of the $(\delta
m^2,\sin^2\theta_{12},\sin^2\theta_{13})$ parameters. The
$n$-sigma bounds on each parameter (the others being
unconstrained) correspond to $\Delta\chi^2=n^2$. The  solid curves
show the results of our previous analysis (see Fig.~9 in
\protect\cite{Ours}). The dotted curves show the effect of the
$\Delta m^2$ estimate in Eq.~(\protect\ref{Deltam2new}), which
includes the revised SK atmospheric neutrino results from
\protect\cite{EPSC}. No significant change is seen in the fit to
$\delta m^2$ and $\sin^2\theta_{12}$. Conversely, the fit to the
$\sin^2\theta_{13}$ parameter becomes less constraining, the
$3\sigma$ upper bound being increased from 0.05 to 0.067.}
\end{figure}
%---------------------------------------------------------------------------

\end{document}